\documentclass[twocolumn,showpacs,amsmath,amssymb,prl]{revtex4}

\usepackage{graphicx}% Include figure files

\begin{document}
\title{A Worm Algorithm for Two-Dimensional Spin Glasses}

\author{Jian-Sheng Wang} 
\affiliation{Singapore-MIT Alliance and
Department of Computational Science, National University of Singapore,
Singapore 117543, Republic of Singapore}

\date{5 June 2005}

\begin{abstract}
A worm algorithm is proposed for the two-dimensional spin glasses. The
method is based on a low-temperature expansion of the partition
function.  The low-temperature configurations of the spin glass on
square lattice can be viewed as strings connecting pairs of frustrated
plaquettes. The worm algorithm directly manipulates these strings.  It
is shown that the worm algorithm is as efficient as any other types of
cluster or replica-exchange algorithms.  The worm algorithm is even
more efficient if free boundary conditions are used.  We obtain
accurate low-temperature specific heat data consistent with a form $ c
\sim T^{-2} \exp\bigl(-2J/(k_BT)\bigr)$, where $T$ is temperature and
$J$ is coupling constant, for the $\pm J$ two-dimensional spin glass.
\end{abstract}

\maketitle

Spin glasses have been studied for many years from various points of
views \cite{Edwards}.  However, the nature of low-temperature phases
is still not clarified.  Much of the work on spin glasses relies on
computer simulations.  Monte Carlo simulation has been one of the main
tools.  The Metropolis algorithm proposed more than half a century ago
served well for many of the simulational work, but for spin glasses,
it is hampered by extremely long relaxation times at low temperatures.
There have been a number of more efficient techniques, noticeably the
simulated tempering \cite{Marinari} and parallel tempering
\cite{exchange}.  In two-dimensions (2D), cluster algorithms
\cite{SW,replicaMC,Liang,Houdayer} exist which are quite efficient.

The idea of generating random walks of loops in Monte Carlo simulation
has a surprisingly long history \cite{earlyloop}.  The loop algorithms
for quantum systems \cite{evertz} are examples.  Recently, several
worm algorithms are proposed \cite{Prokofev,Hitchcock} for the
ferromagnetic Ising and other models.  Such algorithms have the
advantage that the Monte Carlo updates are purely local, while their
effects are global.  The classical worm algorithms based on
high-temperature expansion variables such as $\tanh(J_{ij}/k_{B}T)$ do
not work for spin glasses as the weights would be negative for the
antiferromagnetic interactions. In this paper, we propose a worm
algorithm for the two-dimensional spin glasses.  Our starting point
can be thought of as a low-temperature expansion for the partition
function.  Our algorithm turns out as efficient as the best cluster
algorithms \cite{Houdayer}, replica Monte Carlo \cite{replicaMC} or
replica exchange algorithms \cite{exchange}.  Moreover, large systems
can be simulated since we may only simulate one system at a time.

In the following, we outline a base algorithm and then show how to
enhance it by multi-step moves.  We discuss the efficiency of few
variations of the algorithm.  We report extensive simulation results
for the low-temperature specific heat with periodic and free boundary
conditions.

The weight of a spin-glass configuration is proportional to the
Boltzmann factor $\exp\bigl(\sum_{\langle i\,j\rangle}J_{ij} \sigma_i
\sigma_j/(k_BT)\bigr)$, where $\sigma_i=\pm 1$, and the site $i$ is on an
$L\times L$ square lattice with periodic boundary conditions.  The
summation is over the nearest neighbor pairs.  For simplicity, we
consider the $\pm J$ spin glass where $J_{ij} = +J$ or $-J$ with equal
probability, although the method is not limited to this model.  By
multiplying the weight by a configuration-independent constant, we can
rewrite it in an equivalent form:
\begin{equation}
 \prod_{\langle i\,j\rangle} w^{b_{ij}}, \label{eq-weight}
\end{equation}
where, $w=\exp(-2K)$, $K=J/(k_BT)$; the variable $b_{ij} = \frac{1}{2}
(1-J_{ij}\sigma_i\sigma_j/J)$ represents presence (1) or absence (0)
of an unsatisfied bond.  The bonds live on the dual square lattice.
Note that the variables $b_{ij}$ are not independent.  They should be
set up in such a way that an even number of bonds incident on an
unfrustrated plaquette, while an odd number of bonds incident on a
frustrated plaquette. A groundstate is one such that all the frustrated
plaquettes are paired and connected by strings with minimum total
length \cite{optimize}.  At excited states, closed loops of strings
can form.

The worm algorithm directly manipulates these strings.  The weight,
Eq.~(\ref{eq-weight}), can be sampled with a ``worm'' if we extend the
phase space to include a path of a worm, with a moving head at
location $i$, and a tail at a fixed location $i_0$.  The weight is
exactly the same as before, except that the parity requirement for the
head and tail is reversed.  That is, a frustrated plaquette at head or
tail requires an even number of bonds, and unfrustrated plaquette an
odd number of bonds. The movement of the worm must preserve the
constraint on the bonds.  A valid configuration of the original
problem is formed when the worm traces out a closed loop.

The sites and bonds in the following refer to the dual lattice.  The
base algorithm of the worm movement with a periodic boundary condition
is as follows:

{
\def\myitem#1{\hangindent=2\parindent\hangafter=-5\noindent\llap{#1\enspace\ignorespaces}}
\myitem{1.}Pick a site $i_0$ at random as the starting point.  Set $i
\leftarrow i_0$.

\myitem{2.}Pick a nearest neighbor $j$ with equal probability, and
move it there with probability $w^{1-b_{ij}}$. If it is accepted, flip
the bond variable $b_{ij}$ ($0 \leftrightarrow 1$) along the way,
update $i \leftarrow j$.

\myitem{3.}If $i$ is at the same site as $i_0$ and winding numbers are
even, one Monte Carlo loop is finished (exit and take statistics),
else go to step 2.

}

\noindent We define the winding numbers as the algebraic sum of
displacements $\sum \delta{\bf r}$ divided by the linear size $L$ when
the head and tail meet.  The requirement that the winding numbers must
be even is due to the constraints of the bonds on the dual lattice and
the spins on the original lattice.  A one-to-two mapping to spin
configurations is possible and valid from a bond configuration only if
the worm winds the system an even number of times in both directions.
However, for free boundary conditions, where the spins at the boundary
have fewer neighbors, the winding number constraints are not needed.
Because both the bond variables and spin variables are available, any
desired thermodynamic quantities, such as spin-glass susceptibility,
can be obtained.  The algorithm is nothing but a Metropolis sampling
on the extended phase space.  It is ergodic, any state can be reached
with nonzero probability, except at zero temperature.  We also note
that the difference between a ferromagnetic Ising model and a spin
glass is only at the initial configuration.  In the case of the 2D
ferromagnetic Ising model, because of duality, the same bond
configuration can also be interpreted as a high-temperature expansion
loop configuration at its dual temperature.  Thus, the present
algorithm is exactly a dual algorithm of Ref.~\cite{Prokofev} for the
2D ferromagnetic Ising model.

The basic algorithm can be systematically enhanced by applying the
N-fold way \cite{Bortz} or ``absorbing Markov chain'' method
\cite{Novotny}.  Note that we are only interested in the configuration
when the worm forms a closed path, and do not care about the (Monte
Carlo) dynamics while the worm is making its way to meet the tail.  We
can apply an $n$-step acceleration if it is $n$-step away from the
tail $i_0$.

Consider a set $A$ of states in state space, consisting of the current
state and a collection of neighborhood states of the current state.
For example, in our application, we can consider all the states
reachable from current state in $n-1$ steps of moves or less.  We calculate
the escape probability, $P(\nu|\mu)$, of exiting $A$ to state $\nu$
given that it is in state $\mu$.  Let $W^{AA}_{\mu\nu}$ be the
one-step transition matrix elements of the Markov chain for states
within the set $A$ from $\mu$ to $\nu$, and $W^{A\bar{A}}_{\mu\nu}$
for one-step transition probability with $\mu \in A$, but $\nu
\in\bar{A}$, where $\bar{A}$ is the complement of $A$.  Then the
escape probability is given by
\begin{equation}
P(\nu|\mu) = \sum_{k=0}^\infty [(W^{AA})^k W^{A\bar{A}}]_{\mu\nu} = 
          [(I-W^{AA})^{-1} W^{A\bar{A}}]_{\mu\nu},
\end{equation}
where $I$ is an identity matrix.  The total escape probability is one,
$\sum_{\nu} P(\nu|\mu) = 1$.  Given the current state $\mu$, the state
$\nu \in \bar{A}$ is sampled with the probability $P(\nu|\mu)$.  The
escape probability can be calculated explicitly.  Let the current
state be called 0, and its 4 neighbor states by one step move 1 to 4.
A new state is uniquely specified by a path of the moves defined by
the base algorithm, given the current state.  One-step escape
probability is $P(0\to \mu) = d_0 W_{0\mu}$, $\mu=1,2,3,4$.  $d_0 =
1/( 1-W_{00})$ is fixed by normalization.  This is just the original
N-fold way of Bortz {\sl et al.}~\cite{Bortz}.  For a two-step move,
from the current site 0 to an intermediate site $a$ and reaching $\nu$,
the escape probability is $P(0\to a \to \nu) = d_0 W_{0a}
W_{a\nu}/(1-W_{aa})$ where $d_0$ is again fixed by normalization.
Probabilities for three or more steps are slightly more complicated,
but can be worked out.  In this work, we consider $n=0$ (no
acceleration), and $n=1$ to 4 step accelerations.

At very low temperatures, it can take an exceedingly long time to
generate one loop.  In this case, it is actually correct to interrupt
the simulation by setting a fixed upper limit to the number of steps used
for each loop.  Those attempts that exceed the upper limit will be
treated as rejected moves.

We measure the performance of the algorithm by its correlation times.
The correlation times are defined through the correlation functions of
the overlapping spin-glass order parameter:
\begin{equation}
f(t) = \left[ { \langle Q(t+t') Q(t') \rangle - \langle Q(t') \rangle^2 \over
                \langle Q(t')^2 \rangle - \langle Q(t') \rangle^2 }
  \right]_J,
\end{equation}
where the angular brackets denote average of Monte Carlo loop moves
$t'$ and the outer square brackets mean average over the quenched
random couplings $J_{ij}$.  The quantity $Q$ is the overlap of the
spins of two independent configurations,
\begin{equation}
Q = \bigl| \sum_{i} \sigma_{i}^1 \sigma_{i}^2\bigr|.
\end{equation}

%% Figure 1
\begin{figure}
\includegraphics[width=\columnwidth]{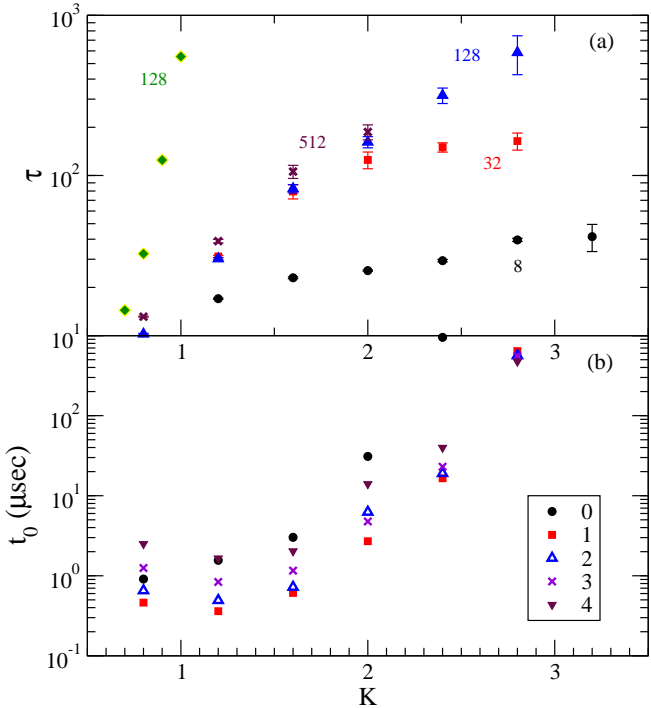}
\caption{\label{fig:Tau} (a) Exponential relaxation times in units of
loop trials of the worm algorithm for system sizes
$L \times L$ where $L=8$, 32, 128, and 512, vs.~dimensionless inverse
temperature $K=J/(k_BT)$.  For comparison, the result of
single-spin-flip Metropolis algorithm on $L=128$ is also plotted.  (b)
CPU time ($\mu$sec) per loop trial per lattice site for a $32 \times
32$ lattice on a $2.4\,$GHz AMD Opteron computer, without (0) or with
$n$-step acceleration. }
\end{figure}

Figure~\ref{fig:Tau} demonstrates the efficiency of the algorithm for
various sizes and inverse dimensionless temperature $K=J/(k_BT)$.  The
results are from linear fits of the form $\ln f(t) = -t/\tau + c$ in a
window $[\tau, 3\tau]$.  The correlation functions are very close to a
pure exponential with $c \approx 0$.  The top part,
Fig.~\ref{fig:Tau}(a), shows correlation time $\tau$ in units of loop
moves, the bottom part (b) shows the central processor unit (CPU)
times.  It is useful to separate the effect of the intrinsic dynamics
defined by the base algorithm from the speed-up due to differences in
detailed implementation.  The correlation time in
Fig.~\ref{fig:Tau}(a) is independent which of the N-fold-way is used
since the N-fold-way preserves the dynamics.  They all give the same
correlation time in units of number of loops generated.  The
bottom part Fig.~\ref{fig:Tau}(b) gives the actual CPU time, $t_0$, in
microsecond for one loop generation divided by the number of spins. This number is only slightly
dependent on system size $L$.  The overall efficiency should be
measured by the product of the two.  It is interesting to note that as
$K$ increases, the correlation time $\tau$ saturates and becomes $K$
independent.  Unfortunately, the time it takes for generating one loop
increases exponentially. Comparing different versions of N-fold-way
acceleration, we found that there is a big improvement going from the
base algorithm to one-step N-fold-way.  Further increasing the step
size does not lead to big improvement until very low temperatures.  It
is clear that, at $T=0$, one step or two step N-fold-way will not be
ergodic, the system can be trapped in a configuration.  However, if we
allow for sufficiently long-ranged multi-step attempts, we can still
make moves even at $T=0$.

%% Figure 2
\begin{figure}
\includegraphics[width=\columnwidth]{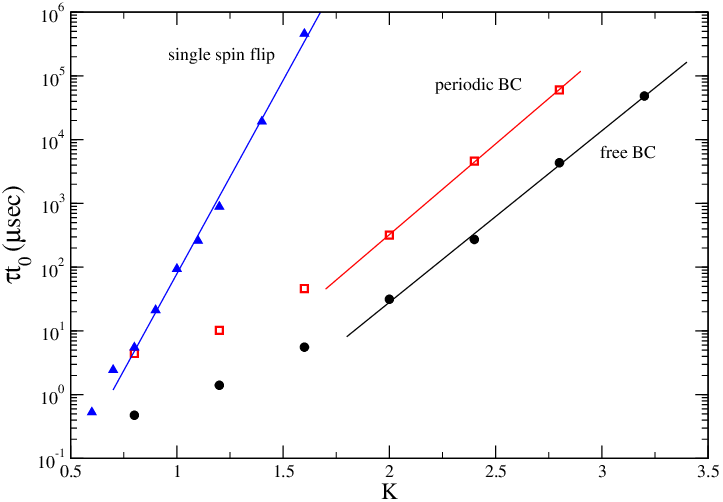}
\caption{\label{fig:sec} Correlation times, $\tau t_0$, measured in
actual CPU times for single-spin-flip, 1-step N-fold-way worm
algorithm with periodic boundary condition and with a free boundary
condition, for $L=128$. The slopes $a$ of the straight line fits to
$\exp(aK)$ are 14.0, 6.5, and 6.2, respectively.}
\end{figure}

What is more remarkable is that a free boundary condition will make
the algorithm even more efficient.  This is because we can always
start from the boundary, or ``outside'' the system, and the worm will
cut the system into pieces.  The worm has a much easier job
terminating as there are order $L$ more possibilities to hit the
boundary comparing to a single site.  Conceptually, we can view the
system as a planar graph.  Various ways of exiting the system can be
regarded as moving to the single dual site representing the
outside. Fig.~\ref{fig:sec} gives the correlation times measured in
real CPU time, the quantity $\tau t_0$.  The meaning of this quantity
is the amount of CPU time needed in order to decorrelate the system
such that the correlation function is reduced to about $e^{-1}$.  This
is a fair comparison between completely different algorithms, but the
results depend on the detail implementation of the computer programs.
From this plot, we see that the free boundary condition case is about
20 times more efficient than periodic boundary conditions.  The
correlation length $\xi$ of the 2D $\pm J$ spin glass diverges as
$\exp(2K)$ \cite{Kardar,Houdayer,Katzgraber}.  If we define the
dynamical critical exponent $z$ as $\tau t_0 \sim \xi^z$, we have $z
\approx 7.0$ and 3.2, for the Metropolis single spin flip and the worm
algorithms, respectively.

We now turn to the specific heat of the spin glass at low
temperatures.  In 1988, Wang and Swendsen found by their replica Monte
Carlo algorithm that the specific heat of the 2D $\pm J$ spin glass
approaches zero according to $c\sim K^{2} \exp(-2K)$
\cite{replicaMCPRB}.  Since the energy gap from the ground states to
the first excited states is $4J$, we might expect that the specific
heat should go as $\exp(-4K)$. Wang and Swendsen gave an argument in
analogous to the 1D Ising model with periodic boundary condition,
where although the minimum excitation is also $4J$, the configuration
appears in the form of a pair of kinks, each one of them can move
freely, so only a single kink with energy $2J$ should be considered
``elementary.''  Indeed, for 1D Ising model, the specific heat goes as
$\exp(-2K)$ in the thermodynamic limit.

However, the above interpretation has been challenged in
Ref.~\cite{Kardar} with a method of exact calculation of the partition
function and recently in \cite{Katzgraber} by Monte Carlo simulation,
but supported in \cite{exact-partition}.  Thus, the problem is still
controversial.

%% Figure 3
\begin{figure}
\includegraphics[width=\columnwidth]{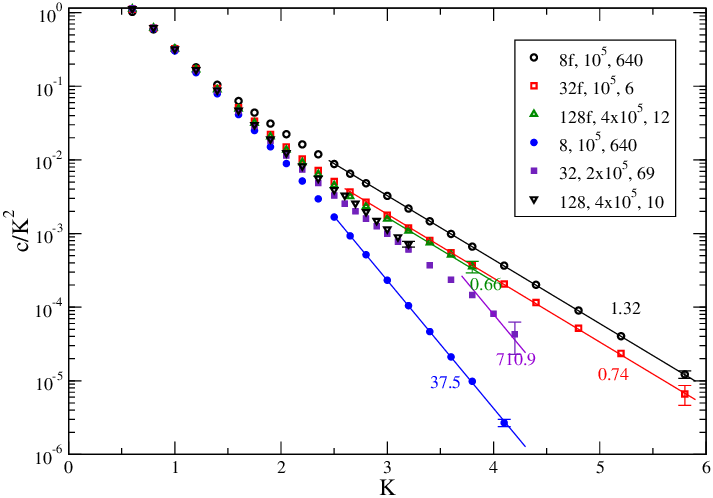}
\caption{\label{fig:lnck} 2D $\pm J$ spin-glass reduced specific heat
$c/K^2$ vs.~$K$.  The legend specifies the lattice size $L$ ($f$ for
free boundary conditions), Monte Carlo steps, and number of samples of
random couplings. The straight lines are fit to $c/K^2 \approx b
\exp(-2K)$ or $b \exp(-4K)$, with $b$ indicated on the lines. The
errors at the largest $K$ are mainly due to sample fluctuations.}
\end{figure}

The present algorithms are well-suited to simulate spin glasses at low
temperatures, particularly the free boundary condition version.
However, as we can see from Fig.~\ref{fig:sec}, the worm algorithms
still have difficulty equilibrating the system when coupling $K > 3$.
Therefore, we have used a combination of several different algorithms.
Each Monte Carlo step consists of several worm loop moves, one sweep
of Metropolis single spin flip, replica Monte Carlo between systems at
neighboring temperatures and at the same temperature.  Three replicas
for each temperature are used.  With these algorithms, we were able to
equilibrate the system down to $K\approx 6$ for $L=8$ and 32, or
$K\approx 4$ for $L=128$.

In Fig.~\ref{fig:lnck}, we present the reduced specific heat $cK^{-2}$
vs.~$K$.  The asymptotic slopes for large $K$ should resolve the issue
of $2J$ vs.~$4J$ controversy.  As can be seen from the figure, with
periodic boundary conditions, the specific heat eventually settles to
$\exp(-4K)$ because of the gap of excitations.  However, as $L$
increases, the crossover to a fast decay appears at a larger and
larger value of $K$. According to Ref.~\cite{exact-partition}, the
cross-over length scale is $l \sim \exp(K)$. This
``finite-size-effect'' length scale appears different from the
correlation length $\xi$, which goes as $\exp(2K)$.  On the other
hand, the free boundary condition results have a rather different
finite-size dependence.  We see that the asymptotic form of
$\exp(-2K)$ is approached much faster in this case.  Of course, since
the energy gap with free boundary conditions is $2J$, the final form
must be $\exp(-2K)$ for finite sizes.  Our data indicate that this is
also the form in the thermodynamic limit.  Some weak size dependence
on the amplitudes is seen, but surprisingly, the thermodynamic limit
is obtained much faster with free boundary conditions than with
periodic boundary conditions.

In summary, we have shown numerically that the low-temperature
specific heat goes like $K^{2} \exp(-2K)$ in the thermodynamic limit.
This is observed much clearer with free boundary conditions.  Our new
calculations support the original argument of Wang and Swendsen
regarding the asymptotic form of heat capacity for 2D $\pm J$ spin
glass.  We demonstrate these results with an efficient worm algorithm.
The worm algorithm presented here should be applicable to any models
defined on a (planar) graph where the concept of dual graph can be
defined.  It does not seem possible for 3D lattices.  It is also
interesting to study the clusters generated in the worm algorithms and
to relate them to other quantities.  More detailed study will be
presented elsewhere.

This work is supported in part by a Faculty Research Grant of National
University of Singapore.

\end{document}